\documentclass[proceedings]{DUBNA} 
\usepackage{epsfig}    
\usepackage{clpsref}    

\title{Heavy-Meson Observables\\ via Dyson-Schwinger Equations}

\author{M.A.\ Ivanov\\Bogoliubov Laboratory of Theoretical Physics,\\
Joint Institute for Nuclear Research, 141980 Dubna, Russia\\[1ex]
\hskip-1.0em{\normalsize \bfseries \sffamily Yu.L.\ Kalinovsky}\\
Laboratory of Computing Techniques and Automation,\\Joint
Institute for Nuclear Research, 141980 Dubna, Russia\\[1ex]
\hskip-1.0em{\normalsize \bfseries \sffamily C.D.\ Roberts}\\
Physics Division, Argonne National Laboratory, Argonne IL 60439-4843, USA.\\
}

\conference{Heavy Quark Physics 5, Dubna, Russia, 6-8 April 2000}

\abstract{We summarise a Dyson-Schwinger-equation-based calculation of an
extensive range of light- and heavy-meson observables, characterised by
heavy-meson leptonic decays, semileptonic heavy-to-heavy and heavy-to-light
transitions - $B \to D^\ast$, $D$, $\rho$, $\pi$; $D \to K^\ast$, $K$, $\pi$,
radiative and strong decays - $B_{(s)}^\ast \to B_{(s)}\gamma$; $D_{(s)}^*\to
D_{(s)}\gamma$, $D \pi$, and the rare $B\to K^\ast \gamma$ flavour-changing
neutral-current process.  In the calculation the heavy-quark mass functions
are approximated by constants, interpreted as their constituent-mass: $\hat
M_c=1.32\,$GeV and $\hat M_b=4.65\,$GeV.}

\keywords{Dyson-Schwinger Equations, Hadron Physics, Heavy-Quark Observables}

\newcommand{\lsim}{\mathrel{\rlap{\lower4pt\hbox{\hskip0pt$\sim$}}
\raise1pt\hbox{$<$}}}

\begin{document} 

\hyphenation{ap-pro-xi-ma-ted}
\hyphenation{mo-men-tum} 

\section{Introduction}
%
The Dyson-Schwinger equations (DSEs)$\,$\cite{RW} provide a nonperturbative,
Poin\-car\'e-covariant, field theoretical approach to the calculation of
hadronic matrix elements, and they have been wi\-de\-ly applied to the
phenomena of continuum strong QCD \cite{bastirev}.  Many applications have
focused on nonhadro\-nic electro\-weak interactions because the electro\-weak
probe is well understood and the interactions therefore explore the structure
of the hadro\-nic target.  These are just the phenomena of interest to this
community.

\FIGURE[hbt]{\epsfig{figure=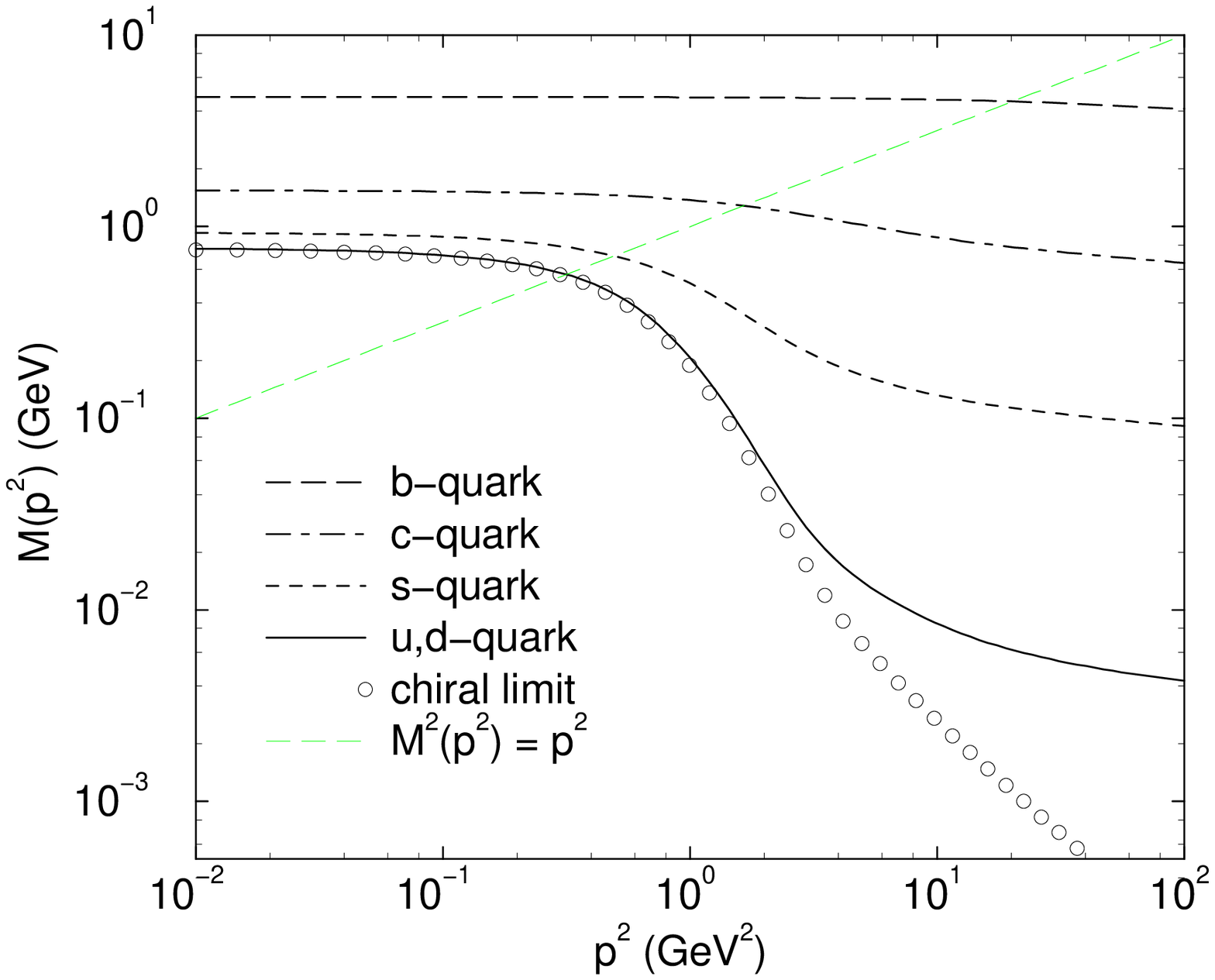,height=5.8cm}\caption{Quark mass function
obtained in Ref.~\protect\cite{mr97} as a solution of the QCD gap equation
with current-quark masses, fixed at a renormalisation point $\zeta= 19\,$GeV:
$m_{u,d}^\zeta = 3.7\,$MeV, $m_s^\zeta = 82\,$MeV, $m_c^\zeta=0.58\,$GeV and
$m_b^\zeta=3.8\,$GeV.  The indicated solutions of $M^2(p^2)=p^2$ define the
Euclidean constituent-quark mass, $M^E_f$, which takes the values:
$M^E_u=0.56\,$GeV, $M^E_s=0.70\,$GeV, $M^E_c= 1.3\,$GeV, $M^E_b= 4.6\,$GeV.
\label{mp2fig}(Figure adapted from Ref.~\protect\cite{IKR}.)}}
In Refs.~\cite{usPLB,IKMR,pmrostock,IKR} both light- {\it and} heavy-mesons
are represented as bound states of a dressed-quark and -anti\-quark, with the
quarks' dressing des\-cribed by a DSE: the QCD {\it gap equation}.  The
general form of the solution of this equation is
\begin{equation}
S_f(p)=Z_f(p^2)/[i\gamma\cdot p + M_f(p^2)]\,,
\end{equation}
$f=u,d,s,c,b$ is the flavour label, and extensive studies have
revealed~\cite{pmrostock,IKR} that while the mass function of a light-quark:
$M_{u,d,s}$, is a rapidly-varying function, that of an heavy-quark is almost
mo\-men\-tum-independent: see Fig.~\ref{mp2fig}.

The behaviour of $M_{c,b}$ (and also that of $Z_{c,b}$, which is not
illustrated here) suggests that the heavy-quark propagator can be
well-ap\-pro\-xi\-ma\-ted by:
\begin{equation}
\label{dsehq}
S_Q(p) = 1/[i\gamma \cdot p + \hat M_Q]\,,\; Q=c,b\,,
\end{equation}
where $\hat M_Q$ is a constituent-heavy-quark mass parameter.  A good
description of observable phenomena requires $\hat M_Q\approx M_Q^E$, see
below.

The dressed-quark propagator can also be written in the form
\begin{equation}
\label{defS}
S_f(p) = -i\gamma\cdot p\,\sigma^f_V(p^2) + \sigma^f_S(p^2)\,.
\end{equation}
and the contrasting, significant mo\-men\-tum-de\-pen\-dence of the
light-quark propagator is efficaciously represented via the algebraic
parametrisation introduced in Ref.~\cite{Rob}
\begin{eqnarray}
\label{SSM}
\bar\sigma^f_S(x)   & = & 
        2 \bar m_f {\cal F}(2 (x + \bar m_f^2))\\
&+ & \nonumber {\cal F}(b_1 x) {\cal F}(b_3 x) 
        \left( b^f_0 + b^f_2 {\cal F}(\epsilon x)\right),\\
\label{SVM}
\bar\sigma^f_V(x) & = & \frac{1}{x+\bar m^2}
\left[ 1 - {\cal F}(2 (x+\bar m_f^2))\right],
\end{eqnarray}
$f=u,s$ (isospin symmetry is assumed), ${\cal F}(y)= (1-{\rm e}^{-y})/y$,
$x=p^2/\lambda^2$; $\bar m_f$ = $m_f/\lambda$; and
\begin{equation}
\bar\sigma_S^f(x)  =  \lambda\,\sigma_S^f(p^2),\;
\bar\sigma_V^f(x)  =  \lambda^2\,\sigma_V^f(p^2)\,,
\end{equation}
with $\lambda=0.566\,$GeV a mass scale.  This algebraic form combines the
effects of confinement and DCSB with free-particle behaviour at large
spacelike $p^2$: $\sigma_V(p^2)\sim 1/p^2$ and $\sigma_S(p^2)\sim m/p^2$.
(With $S(p)$ an entire function a sufficient condition for confinement is
satisfied; i.e., the absence of a Leh\-mann representation for coloured
Schwinger functions~\cite{bastirev}.  $\epsilon=10^{-4}$ is introduced in
Eq.~(\ref{SSM}) only to decouple the large- and in\-ter\-me\-diate-$p^2$
domains: it is not a fitting parameter.)

Bound states in quantum field theory are described by Bethe-Salpeter
amplitudes whose mo\-men\-tum-dependence implements the necessary restriction
of the relative momentum of the constituents in a bound state.  This means
that a further, commonly-used approximation to Eq.~(\ref{dsehq})
\begin{equation}
\label{hqf}
S_Q(k+P) = \frac{1}{2}\,\frac{1 - i \gamma\cdot v}{k\cdot v - E_H}
+ {\rm O}\left(\frac{|k|}{\hat M_{Q}},\frac{E_H}{\hat M_{Q}}\right)
\end{equation}
where $P_\mu=: m_H v_\mu=: (\hat M_Q + E_H)v_\mu$, $m_H$ is the hadron's mass
and $k$ is the momentum of the lighter constituent, can only be reliable if
{\it both} the momentum-space width of the Bethe-Salpeter amplitude,
$\omega_H$, and the binding energy, $E_H$, are significantly less than $\hat
M_Q$.

\section{Heavy-Quark Limit}
%
The DSE framework reproduces all the acknowledged consequences of heavy-quark
symmetry.  In addition, one obtains explicit expressions for the physical
observables in the heavy quark limit; e.g., the leptonic decay constants is
given by
\begin{eqnarray}
\nonumber f_P& = &f_V= \frac{\kappa_f}{\surd m_H}\,\frac{N_c}{2\sqrt{2}
\pi^2}\, \int\limits_0^\infty du \left(\sqrt{ u} - E_H\right)
\nonumber\\ &\times& \varphi_H(z)\left\{ \sigma_S^f(z) +
\frac{1}{2}\sqrt{u}\,\sigma_V^f(z)\right\}\,,
\label{lept}\\
\frac{1}{\kappa_f^2} &=&
\frac{N_c}{4\pi^2}\,
\int\limits_0^\infty du \varphi^2_H(z) 
\left\{ \sigma_S^f(z) + \sqrt{u}\,\sigma_V^f(z)\right\}\,,
\nonumber
\end{eqnarray}
where $z=u-2 E_H \sqrt{ u}$, $f$ labels the meson's lighter quark and
$\varphi_H(z)$ is the scalar function characterising the dominant
Dirac-covariant in the heavy-meson's Bethe-Salpeter amplitude; e.g, the
$\gamma_5$-term for the pseudoscalar and $\gamma_\mu$-term for the vector
meson.

As another example, the semileptonic heavy-to-heavy pseudoscalar transition
form factors ($P_1$ $\to$ $P_2 \ell \nu$) acquire a particularly simple form in
the heavy-quark symmetry limit:
\begin{eqnarray}
\label{fxi}
f_\pm(t)& := & \frac{m_{P_2} \pm 
m_{P_1}}{2\sqrt{m_{P_2}m_{P_1}}}\,\xi_f(w),\\
\label{xif}
\xi_f(w)  & = & \kappa_f^2\,\frac{N_c}{4\pi^2}
\int\limits_0^1 \frac{\tau}{W}
\int\limits_0^\infty du \, \varphi^2_H(z_W) \\
&& \nonumber \times 
\left[\sigma_S^f(z_W) + \sqrt{\frac{u}{W}} \sigma_V^f(z_W)\right]\,,
\end{eqnarray}
with $W= 1 + 2 \tau (1-\tau)(w-1)$, $z_W= u - 2 E_H \sqrt{u/W}$ and
\begin{equation}
w = \frac{m_{P_1}^2 + m_{P_2}^2 - t}{2 m_{P_1} m_{P_2}} = - v_{P_1} \cdot
v_{P_2}\,. 
\end{equation}
The canonical normalisation of the Bethe-Salpeter amplitude automatically
ensures that
\begin{equation}
\label{xione}
\xi_f(w=1) = 1
\end{equation}
and it follows~\cite{usPLB} from Eq.~(\ref{xif}) that
\begin{equation}
\rho^2 := -\left.\frac{d\xi_f}{dw}\right|_{w=1} \geq \frac{1}{3}\,.
\end{equation}

Similar analysis for the heavy-to-heavy transitions with vector mesons in the
final state and for heavy-to-light transitions yields relations between the
form factors that coincide with those observed in Ref.~\cite{IW}; i.e., in
the heavy-quark limit these form factors too are expressible solely in terms
of $\xi_f(w)$.
  
\TABLE[ht]{\caption{The 16 dimension-GeV quantities used in $\chi^2$-fitting
the model parameters.  The values in the ``Obs.'' column are taken from
Refs.~\protect\cite{mr97,PDG,FS}.  (Table adapted from
Ref.~\protect\cite{IKR}.)\label{tableC}}
\begin{tabular}{lll|lll}
        & Obs.  & Calc. & & Obs.  & Calc. \\\hline
$f_\pi$   & 0.131 & 0.146 & $m_\pi$   & 0.138 & 0.130 \\
$f_K  $   & 0.160 & 0.178 & $m_K$     & 0.496 & 0.449 \\
$\langle \bar u u\rangle^{1/3}$ & 0.241 & 0.220 &
        $\langle \bar s s\rangle^{1/3}$ & 0.227 & 0.199\\
$\langle \bar q q\rangle_\pi^{1/3}$ & 0.245 & 0.255& 
        $\langle \bar q q\rangle_K^{1/3}$ & 0.287 & 0.296\\
$f_\rho$   & 0.216      & 0.163 & 
        $f_{K^\ast}$   & 0.244      & 0.253 \\
$\Gamma_{\rho\pi\pi}$ & 0.151     & 0.118 & 
        $\Gamma_{K^\ast (K\pi)}$  & 0.051     & 0.052 \\
$f_D$   & 0.200 $\pm$ 0.030     & 0.213 & 
        $f_{D_s}$ & 0.251 $\pm$ 0.030     & 0.234 \\
$f_B$   & 0.170 $\pm$ 0.035      & 0.182 & 
$g_{B K^\ast \gamma} \hat M_b$ & 2.03 $\pm$ 0.62 & 2.86 \\\hline
\end{tabular}
}
\TABLE[ht]{
\caption{The 26 dimensionless quantities used in $\chi^2$-fitting the model
parameters.  The values in the ``Obs.'' column are taken from
Refs.~\protect\cite{PDG,richman,pirad,cesr96,latt,gHsHpi}.  The light-meson
electromagnetic form factors are calculated in impulse
approximation~\protect\cite{Rob,mark,mrpion} and $\xi(w)$ is obtained
from $f_+^{B\to D}(t)$ via Eq.~(\protect\ref{fxi}).  (Table adapted from
Ref.~\protect\cite{IKR}.)
\label{tableD} }
\begin{tabular}{lll|lll}
        & Obs.  & Calc. & & Obs.  & Calc. \\\hline
$f_+^{B\to D}(0)$ & 0.73 & 0.58  &
        $f_\pi r_\pi$ & 0.44 $\pm$ 0.004 & 0.44   \\
$F_{\pi\,(3.3\,{\rm GeV}^2)}$ & 0.097 $\pm$  0.019  & 0.077 &
        B$(B\to D^\ast)$ & 0.0453 $\pm$ 0.0032 & 0.052\\
$\rho^2$ &  1.53 $\pm$ 0.36 & 1.84 &
        $\alpha^{B\to D^\ast}$ & 1.25 $\pm$ 0.26 & 0.94 \\
$\xi(1.1)$  & 0.86 $\pm$ 0.03& 0.84 &
        $A_{\rm FB}^{B\to D^\ast}$ & 0.19 $\pm$ 0.031 & 0.24 \\
$\xi(1.2)$  & 0.75 $\pm$ 0.05& 0.72 &
        B$(B\to \pi)$ & (1.8 $\pm$ 0.6)$_{\times 10^{-4}}$  & 2.2 \\
$\xi(1.3)$  & 0.66 $\pm$ 0.06& 0.63 &
        $f^{B\to \pi}_{+\,(14.9\,{\rm GeV}^2)}$ & 0.82 $\pm$ 0.17 & 0.82 \\
$\xi(1.4) $ & 0.59 $\pm$ 0.07& 0.56 &
        $f^{B\to \pi}_{+\,(17.9\,{\rm GeV}^2)}$ & 1.19 $\pm$ 0.28 & 1.00 \\
$\xi(1.5) $ & 0.53 $\pm$ 0.08& 0.50 &
        $f^{B\to \pi}_{+\,(20.9\,{\rm GeV}^2)}$ & 1.89 $\pm$ 0.53 & 1.28 \\
B$(B\to D)$ & 0.020 $\pm$ 0.007 & 0.013 &
        B$(B\to \rho)$ & (2.5 $\pm$ 0.9)$_{\times 10^{-4}}$ & 4.8 \\
B$(D\to K^\ast)$ & 0.047 $\pm$ 0.004  & 0.049 &
        $f_+^{D\to K}(0)$ & 0.73 &  0.61 \\
$\frac{V(0)}{A_1(0)}^{D \to K^\ast}$ & 1.89 $\pm$ 0.25 & 1.74 &
        $f_+^{D\to \pi}(0)$ & 0.73 &  0.67 \\
$\frac{\Gamma_L}{\Gamma_T}^{D \to K^\ast}$ & 1.23 $\pm$ 0.13 & 1.17 &
        $g_{B^\ast B\pi}$ & 23.0 $\pm$ 5.0 & 23.2 \\
$\frac{A_2(0)}{A_1(0)}^{D \to K^\ast}$ & 0.73 $\pm$ 0.15 & 0.87 &
        $g_{D^\ast D\pi}$ & 10.0 $\pm$ 1.3 & 11.0 \\\hline
\end{tabular}
}

\section{Results}
%
This phenomenological application of DSE methods to the calculation of
heavy-quark observables is founded on a large body of work that has focused
on light-meson physics.  References~\cite{bastirev,echaya} provide an
overview.  Herein we summarise results for an extensive but not exhaustive
body of observables: heavy-meson leptonic decays, semileptonic heavy-to-heavy
and heavy-to-light transitions - $B \to D^\ast$, $D$, $\rho$, $\pi$; $D \to
K^\ast$, $K$, $\pi$, radiative and strong decays - $B_{(s)}^\ast \to
B_{(s)}\gamma$; $D_{(s)}^*\to D_{(s)}\gamma$, $D \pi$, and the rare $B\to
K^\ast \gamma$ flavour-chan\-ging neutral-current process.

In Ref.~\cite{IKR} an algebraic characterisation of the dressed-quark
propagators and bound state Bethe-Salpe\-ter amplitudes employing ten
parameters, plus the four quark masses, was used in a $\chi^2$-fit to $N_{\rm
obs}=42$ heavy- and light-meson observables.  That yielded: $\chi^2/{\rm
d.o.f} = 1.75$ and $\chi^2/N_{\rm obs} = 1.17$, and the quality of the fit is
illustrated in Tables~\ref{tableC} and~\ref{tableD}.  Using an approximating
algebraic representation of the functions involved materially simplifies the
calculation of observables, bypassing the repeated solving of nonlinear,
coupled integral equations.  It is a useful but not necessary artefice, as
Refs.~\cite{mr97,mtvector,mtpion,mtkaon} make plain.

The fitting yielded dressed-quark-propagator parameter values
\begin{equation}
\label{tableB} 
\begin{array}{c|lll}
      & \;\bar m_f& b_1^f & b_2^f\\\hline
 u\;  & \;0.00948 & 2.94 & 0.733 \\
 s\;  & \;0.210   & 3.18 & 0.858 
\end{array}\,,
\end{equation}
with $b_0^{u,s}=0.131$, $0.105$, $b_3^{u,s}=0.185$, which were not varied
i.e., $b_{0,3}^{u,s}$ retain the values fixed in previous studies of
light-meson observables~\cite{mark}.  The dimensionless $u,s$ current-quark
masses in Eq.~(\ref{tableB}) correspond to $m_u = 5.4\,$MeV and
$m_s=119\,$MeV, and these algebraic propagators yield Euclidean
constituent-light-quark masses: $M^E_u = 0.36\,$GeV and $M^E_s=0.49\,$GeV.

The fitted constituent-heavy-quark mass parameters are
\begin{equation}
\label{HQmass}
\hat M_c = 1.32\,\mbox{GeV~and}\; \hat M_b=4.65\,\mbox{GeV}\,,
\end{equation}
consistent with the estimates reported in Ref.~\cite{PDG} and hence the
heavy-meson binding energy is large:
\begin{equation}
\label{EB}
\begin{array}{l}
E_D:= m_D - \hat M_c = 0.67\,{\rm GeV}\,, \\
E_B:= m_B - \hat M_b = 0.70\,{\rm GeV}\,.
\end{array}
\end{equation}
These values yield $E_D/\hat M_c= 0.51$ and $E_B/\hat M_b= 0.15$, and provide
an indication that while an heavy-quark expansion, Eq.~(\ref{hqf}), will be
accurate for the $b$-quark it will provide a poor approximation for the
$c$-quark.  The constituent-heavy-quark-masses in Eq.~(\ref{HQmass}), obtained
in a Poincar\'e covariant approach~\cite{IKR}, are, respectively, $\sim 25$\%
and $\sim 10$\% smaller than the values used in nonrelativistic models.

Reference~\cite{IKR} represented the dominant scalar function in the
light-pseudoscalar-meson Bethe-Salpe\-ter amplitude as
\begin{equation} 
\label{piKamp}
{\cal E}_{P}(k^2)  =  \frac{1}{\hat f_{P}}\,B_{P}(k^2)\,,\;P=\pi,K\,;
\end{equation}
constructed via $B_P:=\left.B_u\right|_{b_0^u\to b_0^P}$ using
Eq.~(\ref{defS}) with, e.g., $\hat f_\pi= f_\pi/\surd 2$.  This {\it Ansatz}
follows from the constraints imposed by the axial-vector Ward-Takahashi
identity and the fit yielded
\begin{equation}
b_0^\pi = 0.204\,,\; b_0^K= 0.319\,.
\end{equation}
  
The exploration of light-vector-meson properties is less extensive than that
of light-pseudo\-scalar-mesons, and this is not peculiar to DSE analyses.
Hence, following, e.g., Refs.~\cite{mike}, the algebraic characterisation of
Ref.~\cite{IKR} used
\begin{equation}
\label{gammaV}
\varphi(k^2) = 1/(1+k^4/\omega_V^4)\,;
\end{equation}
i.e., a one-parameter form to describe the dominant scalar function in the
vector Bethe-Salpeter amplitude.  This is merely a simple, efficacious {\it
Ansatz}, which might now be improved by building upon
Refs.~\cite{mtvector,mtpion}.  No reliable Bethe-Salpeter equation studies
exist for heavy-mesons (an efficacious Bethe-Salpeter kernel is still being
sought) and therefore the following single-parameter form was employed to
represent the amplitudes:
\begin{equation}
\varphi_H(k^2) = \exp(-k^2/\omega_H^2)\,.
\end{equation}
The $\chi^2$-fit produced:
\begin{equation}
\begin{array}{c|l}
      & \omega_V^{\rm GeV} \\ \hline
\rho  & 0.515 \\
K^\ast& 0.817
\end{array}\;\;
\begin{array}{c|l}
      & \omega_H^{\rm GeV} \\ \hline
   D  & 1.81 \\
   B  & 1.81
\end{array}\;\;
\end{equation}
The ordering in magnitude is qualitatively understandable: the heavier the
meson the smaller the spacetime volume occupied.  In addition, the result:
$\omega_D = \omega_B$, which means that the Compton wavelength of the
$c$-quark is greater than the length-scale characterising the bound state's
extent, emphasises that Eq.~(\ref{hqf}) must provide a poor approximation for
the $c$-quark.

\FIGURE[ht]{\epsfig{figure=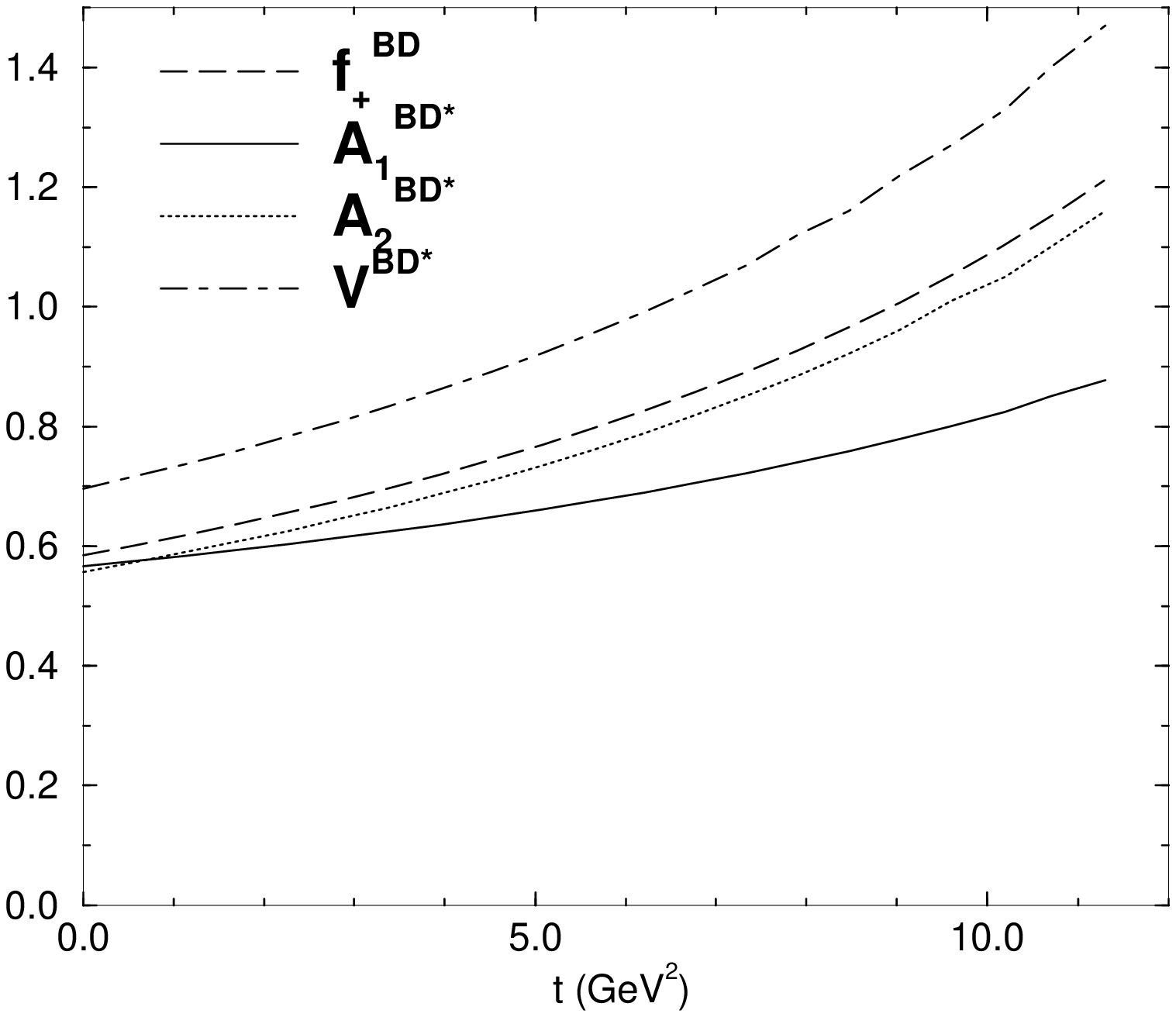,height=5.7cm}\hspace*{3ex}\epsfig{figure=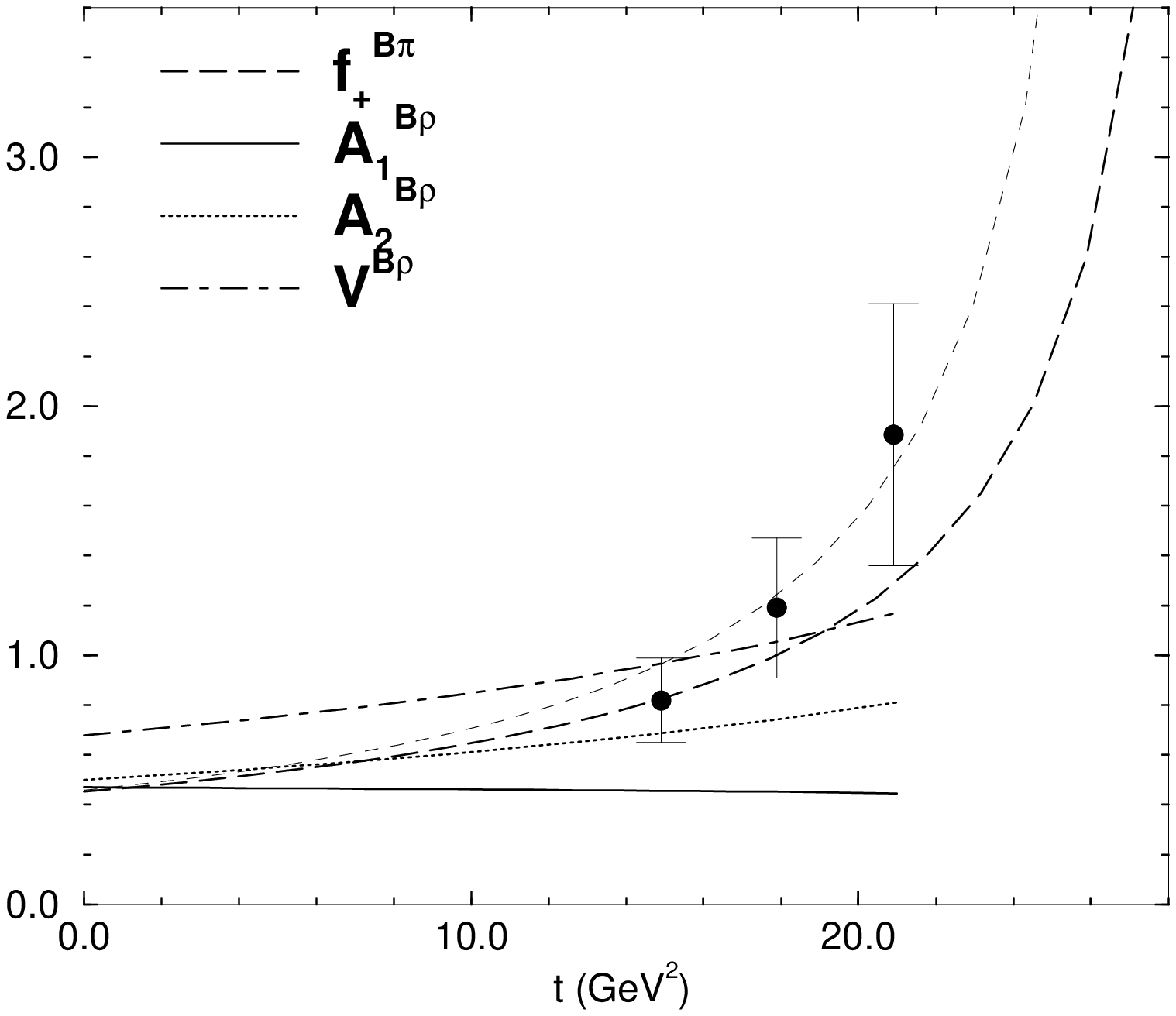,height=5.7cm}
\caption{\label{figa} Left panel: calculated semileptonic $B\to D$ and $B\to
D^\ast$ form factors.  Right panel: the semileptonic $B\to \pi$ and $B\to
\rho$ form factors with, for comparison, data from a lattice
simulation~\protect\cite{latt} and a vector dominance, monopole model:
$f_+^{B\to \pi}(t)= 0.46/(1-t/m_{B^\ast}^2)$, $m_{B^\ast} = 5.325\,$GeV, the
light, short-dashed line.  Monopole fits to the model's results are given in
Eqs.~(\protect\ref{mono}) and (\protect\ref{monomass}).  (Figure adapted from
Ref.~\protect\cite{IKR}.)}
}

\FIGURE[ht]{\epsfig{figure=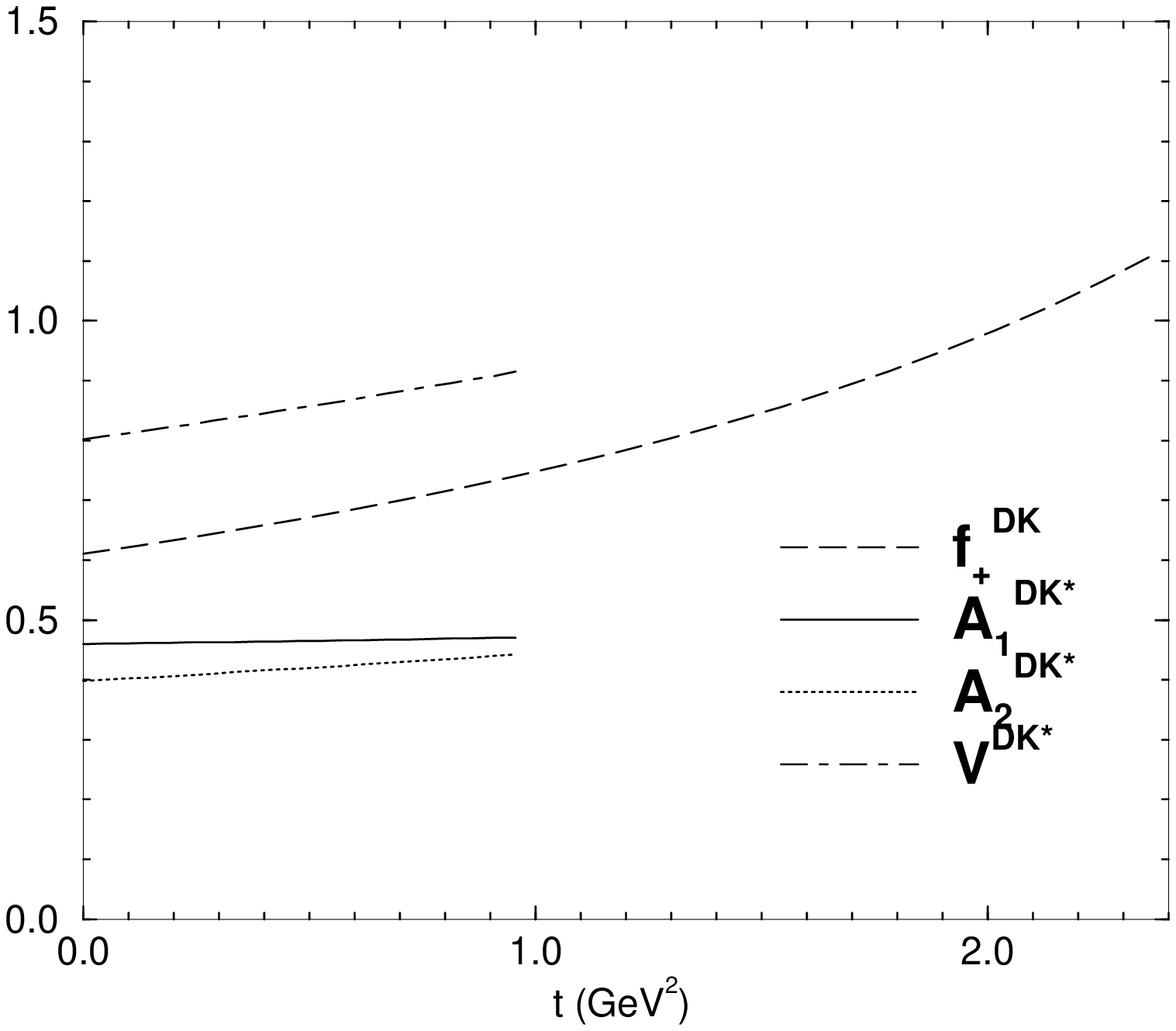,height=5.7cm}\hspace*{3ex}\epsfig{figure=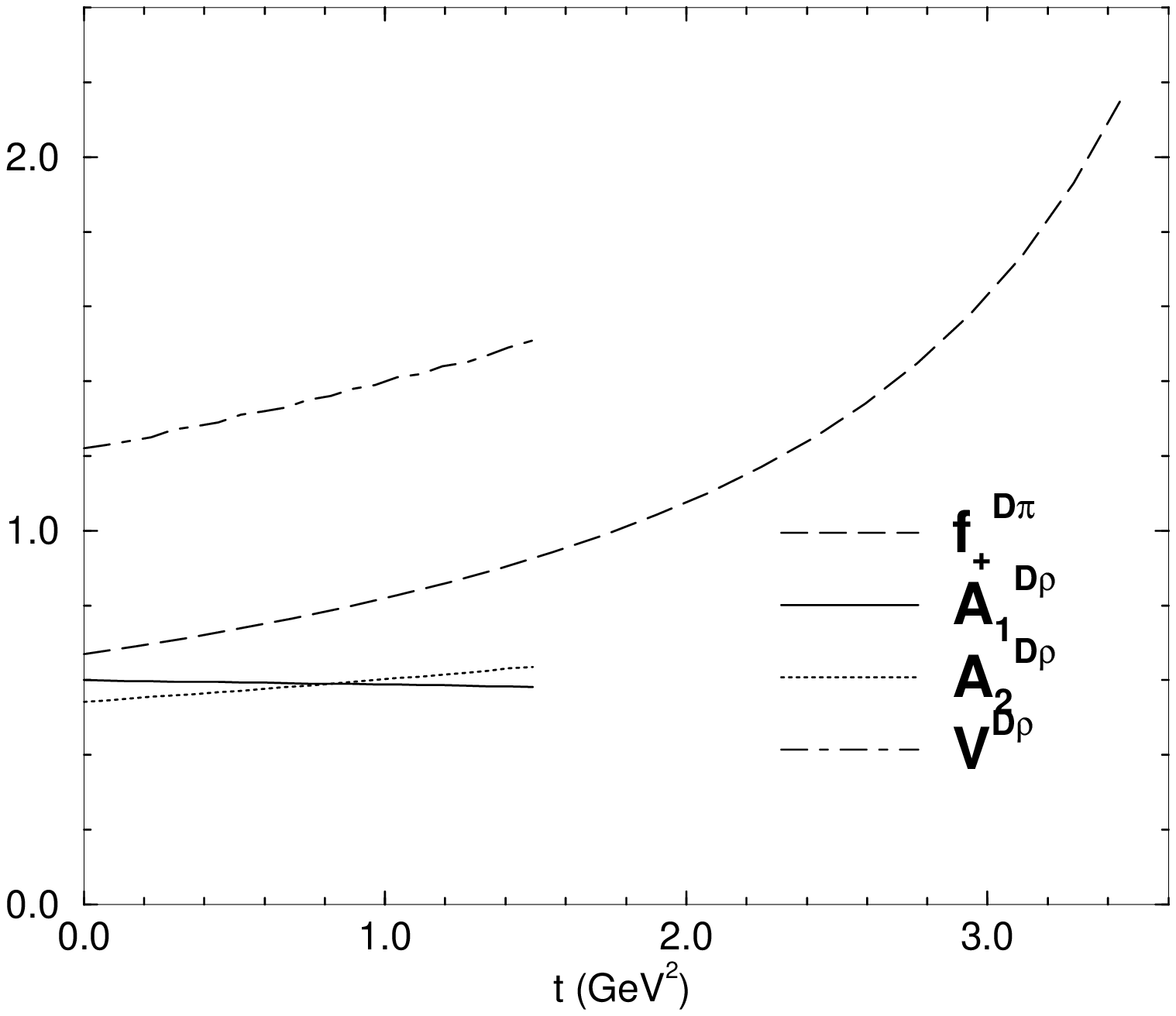,height=5.6cm}
\caption{\label{figb} Calculated semileptonic $D\to K$ and $D\to K^\ast$
(left panel), $D\to \pi$ and $D\to \rho$ (right panel) form factors.
Monopole fits to the calculated results are given in
Eqs.~(\protect\ref{mono}) and (\protect\ref{monomass}).  (Figure adapted from
Ref.~\protect\cite{IKR}.)}  }

With the model's parameters fixed, it is possible to calculate a wide range
of other light- and heavy-meson observables.  Some of the results are
summarised in Tables~\ref{tableE}-\ref{tableG}, while Figs.~\ref{figa}
and~\ref{figb} depict the calculated $t$-depen\-dence of the semileptonic
transition form factors that are the hadronic manifestation of the $b\to c$,
$b\to u$, $c\to s$ and $c\to d$ transitions.  The form factors can be {\it
approximated} by the monopole
\begin{equation}
\label{mono}
h(t) = h(0)/[1 - t/h_1]\,,
\end{equation}
with $h(0)$ given in Tables~\ref{tableF} and~\ref{tableG}, and $h_1$, in
GeV$^2$, listed in Eq.~(\ref{monomass}).
\begin{equation}
\label{monomass}
\begin{array}{l|rrrr}
        & h_1^{f_+} & h_1^{A_1} & h_1^{A_2} & h_1^{V}\\\hline
B\to D, D^\ast& (4.63)^2 & (5.73)^2 & (4.64)^2 & (4.61)^2\\
B\to \pi,\rho & (5.58)^2 & -(21.5)^2 & (6.94)^2 & (7.06)^2\\
D\to K,K^\ast & (2.31)^2 & (6.70)^2 & (3.09)^2 & (2.78)^2 \\
D\to \pi,\rho & (2.25)^2 & -(7.06)^2 & (3.17)^2 & (2.80)^2
\end{array}
\end{equation}

With these calculated results it is possible to check the fidelity of
heavy-quark symmetry limits.  The universal function characterising
semileptonic transitions in the heavy-quark symmetry limit, $\xi(w)$, can be
obtained most reliably from $B\to D,D^\ast$ transitions, if it can be
obtained at all.  Using Eq.~(\ref{fxi}) to extract it from $f_+^{B\to D}(t)$
one obtains
\begin{equation}
\label{xifp}
\xi^{f_+}(1)= 1.08\,,
\end{equation}
which is a measurable deviation from Eq.~(\ref{xione}).  The calculated form
of $\xi^{f_+}(w)/\xi^{f_+}(0)$ is depicted in Fig.~\ref{fige} and compared
with two experimental fits~\cite{cesr96}:
\begin{eqnarray}
\xi(w) & = & 1 - \rho^2\,( w - 1), \\
& & \nonumber  \rho^2 = 0.91\pm 0.15 \pm 0.16\,,
\label{cesrlinear}\\
\xi(w) & = & \frac{2}{w+1}\,\exp\left[(1-2\rho^2) \frac{w-1}{w+1}\right]\!\!,\\
&& \nonumber \rho^2 = 1.53 \pm 0.36 \pm 0.14\,.
\label{cesrnonlinear}
\end{eqnarray}
The calculated result is well approximated by
\begin{equation}
\label{fitIW}
\xi^{f_+}(w) = \frac{1}{1 + \tilde\rho^2_{f_+}\,(w-1)  }\,,
\;\tilde\rho^2_{f_+}=1.98\,.
\end{equation}

\TABLE[ht]{
\caption{Calculated values of a range of observables not included in fitting
the model's parameters.  The ``Obs.''  values are extracted from
Refs.~\protect\cite{PDG,FS,richman,kaondat}.  (Table adapted from
Ref.~\protect\cite{IKR}.)
\label{tableE} }
\begin{tabular}{lll|lll}
        & Obs.  & Calc. & & Obs.  & Calc. \\\hline
$f_K r_K$       &   0.472 $\pm$ 0.038 & 0.46 &     
        $-f_K^2 r_{K^0}^2$ &  (0.19 $\pm$ 0.05)$^2$ & (0.10)$^2$   \\
$g_{\rho\pi\pi}$ & 6.05 $\pm$ 0.02 & 5.27  &
        $\Gamma_{D^{\ast 0} }$ (MeV)& $ < 2.1 $  & 0.020   \\
$g_{K^\ast K \pi^0}$ & 6.41 $\pm$ 0.06 & 5.96  & 
        $\Gamma_{D^{\ast +}}$ (keV) &  $< 131$ & 37.9 \\
$g_\rho$ & 5.03 $\pm$ 0.012 & 5.27  & 
        $\Gamma_{D_s^{\ast} D_s \gamma}$ (MeV)& $< 1.9$  & 0.001    \\
$f_{D^\ast}$ (GeV) &   & 0.290   & 
        $\Gamma_{B^{\ast +} B^+ \gamma}$ (keV)&   & 0.030    \\
$f_{D^\ast_s}$ (GeV)&   & 0.298   & 
        $\Gamma_{B^{\ast 0} B^0 \gamma}$ (keV)&  &  0.015 \\
$f_{B_s}$ (GeV) & 0.195 $\pm$ 0.035 & 0.194  & 
        $\Gamma_{B_s^{\ast} B_s \gamma}$ (keV)&  &  0.011 \\
$f_{B^\ast}$ (GeV)&   & 0.200   &
        B($D^{\ast +}\!\to D^+ \pi^0$) & 0.306 $\pm$ 0.025 & 0.316 \\
$f_{B^\ast_s}$ (GeV)&   & 0.209   & 
        B($D^{\ast +}\!\to D^0 \pi^+$) & 0.683 $\pm$ 0.014 & 0.683 \\
$f_{D_s}/f_D$ & 1.10 $\pm$ 0.06 &  1.10  &
        B($D^{\ast +}\!\to D^+ \gamma$) & 0.011~$^{+ 0.021}_{-0.007}$ & 0.001 \\
$f_{B_s}/f_B$  & 1.14 $\pm$ 0.08  & 1.07   &
        B($D^{\ast 0}\!\to D^0 \pi^0$) &  0.619 $\pm$ 0.029 & 0.826 \\
$f_{D^\ast}/f_D$ &       &  1.36  &
        B($D^{\ast 0}\!\to D^0 \gamma$) &  0.381 $\pm$ 0.029 &  0.174 \\
$f_{B^\ast}/f_B$  &       & 1.10   &
        B($B \to K^\ast \gamma$) & (5.7 $\pm$ 3.3)$_{10^{-5}}$ & 11.4 \\\hline
\end{tabular}
}

\TABLE[ht]{\caption{Calculated values of some $b\to c$ and $b\to u$
transition form factor observables not included in fitting the model's
parameters.  The ``Obs.''  values are extracted from
Refs.~\protect\cite{PDG,richman}.  (Table adapted from
Ref.~\protect\cite{IKR}.)
\label{tableF} }
\begin{tabular}{lll|lll}
        & Obs.  & Calc. & & Obs.  & Calc. \\\hline
$A_1^{B\to D^\ast}(0)$ &   & 0.57   & 
        $A_1^{B\to D^\ast}(t_{\rm max}^{B\to D^\ast})$ &   & 0.88   \\
$A_2^{B\to D^\ast}(0)$ &   & 0.56   & 
        $A_2^{B\to D^\ast}(t_{\rm max}^{B\to D^\ast})$ &   & 1.16   \\
$V^{B\to D^\ast}(0)$ &   & 0.70   & 
        $V^{B\to D^\ast}(t_{\rm max}^{B\to D^\ast})$ &   & 1.47   \\
$R_1^{B\to D^\ast}(1)$ &   1.30 $\pm$ 0.39   & 1.32 & 
        $R_2^{B\to D^\ast}(1)$ & 0.64 $\pm$ 0.29 & 1.04   \\
$R_1^{B\to D^\ast}(w_{\rm max})$ &      & 1.23 & 
        $R_2^{B\to D^\ast}(w_{\rm max})$ &  & 0.98   \\
$\alpha^{B\to\rho}$ &   & 0.60 &
        $f_+^{B\to D}(t_{\rm max}^{B\to D})$ &   & 1.21   \\
$f_+^{B\to \pi}(0)$ & & 0.45 &
        $f_+^{B\to \pi}(t_{\rm max}^{B\to \pi})$ & & 3.73 \\
$A_1^{B\to \rho}(0)$ &   & 0.47   &
        $A_1^{B\to \rho}(t_{\rm max}^{B\to \rho})$ &   & 0.45   \\
$A_2^{B\to \rho}(0)$ &   & 0.50   & 
        $A_2^{B\to \rho}(t_{\rm max}^{B\to \rho})$ &   & 0.81   \\
$V^{B\to \rho}(0)$   &   & 0.68   & 
        $V^{B\to \rho}(t_{\rm max}^{B\to \rho})$ &   & 1.17   \\
$R_1^{B\to \rho}(1)$ &      & 1.15 & 
        $R_2^{B\to \rho}(1)$ &  & 0.80   \\
$R_1^{B\to \rho}(w_{\rm max})$ &      & 1.44 & 
        $R_2^{B\to \rho}(w_{\rm max})$ &  & 1.06   \\\hline
\end{tabular}
}

\TABLE[ht]{\caption{Calculated values of some $c\to s$ and $c\to d$
transition form factor observables not included in fitting the model's
parameters.  The ``Obs.''  values are extracted from
Refs.~\protect\cite{PDG,richman}.  (Table adapted from
Ref.~\protect\cite{IKR}.)
\label{tableG} }
\begin{tabular}{lll|lll}
        & Obs.  & Calc. & & Obs.  & Calc. \\\hline
B($D^+\to \rho^0$) &   & 0.032   &
        $\alpha^{D \to \rho}$ &   & 1.03  \\
${\rm B}(D^0\to K^-)$   &  0.037 $\pm$ 0.002  &  0.036   &
        $\frac{{\rm B}(D\to \rho^0)}
        {{\rm B}(D\to K^\ast)}$ & 0.044 $\pm$ 0.034  & 0.065   \\
$A_1^{D\to K^\ast}(0)$ & 0.56 $\pm$ 0.04 & 0.46   & 
        $A_1^{D\to K^\ast}(t_{\rm max}^{D\to K^\ast})$ &0.66 $\pm$ 0.05 & 0.47 \\
$A_2^{D\to K^\ast}(0)$ & 0.39 $\pm$ 0.08 & 0.40   & 
      $A_2^{D\to K^\ast}(t_{\rm max}^{D\to K^\ast})$ & 0.46 $\pm$ 0.09 & 0.44 \\
$V^{D\to K^\ast}(0)$ & 1.1 $\pm$ 0.2 & 0.80   & 
        $V^{D\to K^\ast}(t_{\rm max}^{D\to K^\ast})$ & 1.4 $\pm$ 0.3 & 0.92 \\
$R_1^{D\to K^\ast}(1)$ &      & 1.72 & 
        $R_2^{D\to K^\ast}(1)$ &  & 0.83   \\
$R_1^{D\to K^\ast}(w_{\rm max})$ &      & 1.74 & 
        $R_2^{D\to K^\ast}(w_{\rm max})$ &  & 0.87   \\
$\frac{{\rm B}(D^0\to \pi)}
        {{\rm B}(D^0\to K)}$ & 0.103 $\pm$ 0.039  & 0.098   &
        $f_+^{D\to K}(t_{\rm max}^{D\to K})$ & 1.31 $\pm$ 0.04 & 1.11 \\
$\frac{f_+^{D\to \pi}(0)}{f_+^{D\to K}(0)}$ & 1.2 $\pm$ 0.3 &  1.10  &
        $f_+^{D\to \pi}(t_{\rm max}^{D\to \pi})$ &  & 2.18 \\
$A_1^{D\to \rho}(0)$ &   & 0.60   &
        $A_1^{D\to \rho}(t_{\rm max}^{D\to \rho})$  && 0.58 \\
$A_2^{D\to \rho}(0)$ &   & 0.54   &
        $A_2^{D\to \rho}(t_{\rm max}^{D\to \rho})$  && 0.64 \\
$V^{D\to \rho}(0)$   &   & 1.22   &
        $V^{D\to \rho}(t_{\rm max}^{D\to \rho})$ & & 1.51 \\
$R_1^{D\to \rho}(1)$ &      & 2.08 & 
        $R_2^{D\to \rho}(1)$ &  & 0.88   \\
$R_1^{D\to \rho}(w_{\rm max})$ &      & 2.03 & 
        $R_2^{D\to \rho}(w_{\rm max})$ &  & 0.91   \\\hline
\end{tabular}
}

\FIGURE[ht]{\epsfig{figure=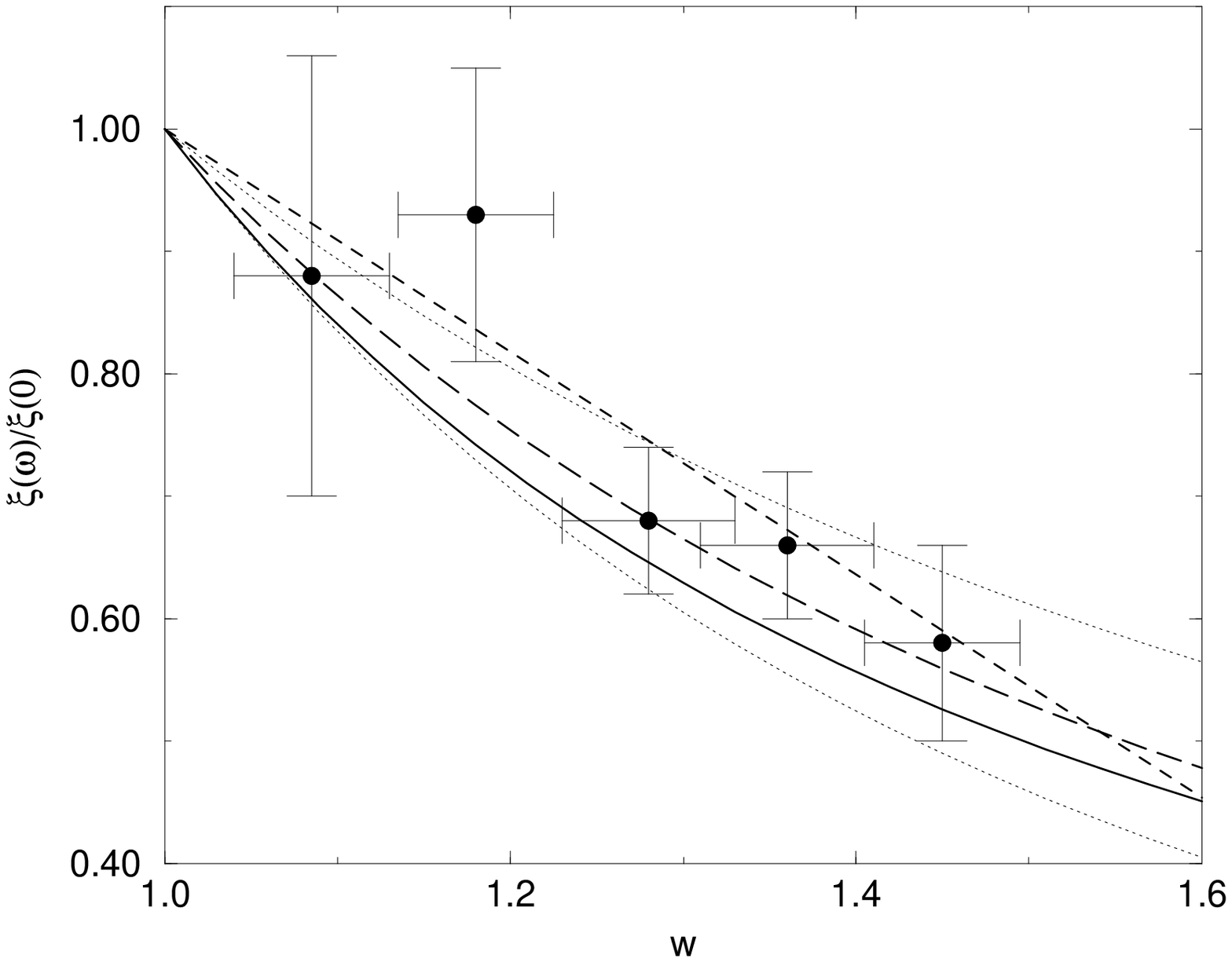,height=5.5cm}
\caption{\label{fige} Calculated form of $\xi(w)$ compared with experimental
analyses.  Model: solid line.  Experiment: short-dashed line, linear fit from
Ref.~\protect\cite{cesr96}, Eq.~(\protect\ref{cesrlinear}); long-dashed line,
nonlinear fit from Ref.~\protect\cite{cesr96},
Eq.~(\protect\ref{cesrnonlinear}).  The two light, dotted lines are this
nonlinear fit evaluated with the extreme values of $\rho^2$: upper line,
$\rho^2= 1.17$ and lower line, $\rho^2=1.89$; data points,
Ref.~\protect\cite{argus93}.  (Figure adapted from Ref.~\protect\cite{IKR}.)}
}

The semileptonic $B\to D^\ast$ transition can also be used to extract
$\xi(w)$.  That yields~\cite{IKR} 
\begin{equation}
\xi^{A_1}(1)= 0.987\,,\;\xi^{A_2}(1)=1.03\,,\;
\xi^{V}(1)= 1.30\,,
\end{equation}
and an $w$-dependence well-described by Eq.~(\ref{fitIW}) but with
\begin{equation}
\tilde\rho^2_{A_1}=1.79\,,\;
\tilde\rho^2_{A_2}=1.99\,,\;
\tilde\rho^2_{V}=2.02\,.
\end{equation}
These and other results in Ref.~\cite{IKR} furnish a measure of the degree to
which heavy-quark symmetry is respected in $b\to c$ processes.  Combining
them it is clear that even in this case, which is the nearest contemporary
realisation of the heavy-quark symmetry limit, corrections of $\lsim 30$\%
must be expected.

\section{Epilogue}
Herein we have summarised a direct extension of DSE-based phenomenology to
experimentally accessible heavy-meson observables~\cite{IKR}.  That extension
explored the fidelity of a simple approximation, Eq.~(\ref{dsehq}), to the
dressed-heavy-quark propagator, and yields a {\it unified} and uniformly
accurate description of an extensive body of light- and heavy-meson
observables.  Algebraic analysis proves~\cite{IKR} that in the heavy-quark
limit pseudoscalar meson masses grow linearly with the mass of their heaviest
constituent; i.e., $m_P \propto \hat m_Q$, while the numerical results
indicate that corrections to the heavy-quark symmetry limit of $\lsim 30$\%
are encountered in $b\to c$ transitions and that these corrections can be as
large as a factor of two in $c\to d$ transitions.

The calculation of the semileptonic transition form factors for $B$- and
$D$-mesons on their entire kinematic domain and with the light-quark sector
well constrained is potentially useful in the experimental extraction of the
CKM matrix elements $V_{cb}$, $V_{ub}$.  That is also true of the calculation
of the leptonic decay constants; e.g., accurate knowledge of $f_B$ can assist
in the determination of $V_{td}$.  The calculations show that the leptonic
decay constants for $D_f$-mesons do not lie on the heavy-quark $1/\surd\hat
m_Q$-trajectory, and provide an estimate of the total width of the
$D_{(s)}^{\ast +}$- and $D^{\ast 0}$-mesons, for which currently there are
only experimental upper-bounds.

The model we have summarised employs simple parametrisations for the
dressed-quark propagators and meson Bethe-Salpeter amplitudes.  Its efficacy
supports the interpretation that heavy- and light-mesons are both simply
finite-size bound states of dressed-quarks and -antiquarks; i.e., they are
not qualitatively different.  Furthermore, this efficacy demonstrates that
the qualitative elements of the Poincar\'e-covariant DSE-framework are rich
enough to account for the gamut of strong interaction phenomena; i.e., as
observed too in Ref.~\cite{a1b1}, no essential element is inherently lacking.

Naturally the model can be improved; e.g., via a wider study of
light-vector-meson observables, so as to more tightly constrain their
properties, perhaps using direct Bethe-Salpeter equation studies like
Refs.~\cite{mtvector,mtpion} as a foundation for improved models of the
vector meson Bethe-Salpeter amplitudes.  Along this path, a more significant
extension is the development of a Ward-Takahashi-identity-preserving
Bethe-Salpeter kernel applicable to the study of heavy-meson masses.  That
would provide further insight into the structure of heavy-meson bound state
amplitudes, an integral part of these calculations for which only rudimentary
models are currently available.  It would also assist in constraining DSE
phe\-no\-me\-no\-lo\-gy via a comparison with calculations and models of the
heavy-quark potential.

\medskip

{\bf Acknowledgments.}~~C.D.R. is grateful for the hospitality and support of
the Erwin Schr\"o\-dinger Institute for Mathematical Physics in Vienna, where
this contribution was completed.  The work of C.D.R.  was supported by the US
Department of Energy, Nuclear Physics Division, under contract
no. W-31-109-ENG-38.


\begin{thebibliography}{99}
%
\bibitem{RW} C.D.~Roberts and A.G.~Williams, Prog.\ Part.\ Nucl.\ Phys.\ {\bf
33} (1994) 477.
%
\bibitem{bastirev} C.D.\ Roberts and S.M.\ Schmidt, ``Dyson-Schwinger
Equations: Density, Temperature and Continuum Strong QCD,'' to appear in
Prog. Part. Nucl. Phys. {\bf 45} (2000).
%
\bibitem{usPLB} M.A.~Ivanov, Yu.L.~Kalinovsky, P.~Maris and C.D.~Roberts,
Phys.\ Lett.\ {\bf B416} (1998) 29.
%
\bibitem{IKMR} M.A.~Ivanov, Yu.L.~Kalinovsky, P.~Maris and C.D.~Roberts,
Phys.\ Rev.\ {\bf C 57} (1998) 1991;
%
\bibitem{pmrostock} P.~Maris and C.D.~Roberts, ``Differences between heavy
and light quarks,'' in Proc. of the {\it IVth International Workshop on
Progress in Heavy Quark Physics}, eds. M.~Beyer, T.~Mannel and H.~Schr\"oder
(University of Rostock, Rostock, 1988) pp.~159-162.
%
\bibitem{IKR} M.A.~Ivanov, Yu.L.~Kalinovsky and C.D.~Roberts, Phys.\ Rev.\
{\bf D 60} (1999) 034018.
%
\bibitem{mr97} P. Maris and C. D. Roberts, Phys. Rev. C {\bf 56}, 3369
(1997); P. Maris, C. D. Roberts and P. C. Tandy, Phys. Lett. B {\bf 420}, 267
(1998).
%
\bibitem{Rob} C.D.~Roberts, Nucl.\ Phys.\ {\bf A 605} (1996) 475.
%
\bibitem{IW} N.~Isgur and M.B.~Wise, Phys.\ Lett.\ {\bf B 232} (1989) 113;
%
{\it ibid} {\bf B 237} (1990) 527.
%
\bibitem{echaya} C.D.~Roberts, Fiz. \'Elem. Chastits At. Yadra {\bf 30},
537 (1999) (Phys. Part. Nucl. {\bf 30}, 223 (1999)).
%
\bibitem{mtvector} P.~Maris and P.~C.~Tandy, Phys.\ Rev.\ {\bf C 60} (1999)
055214.
%
\bibitem{mtpion} P.~Maris and P.~C.~Tandy, Phys.\ Rev.\ {\bf C 61} (2000)
045202.
%
\bibitem{mtkaon} P.~Maris and P.~C.~Tandy, ``The $\pi$, $K^+$ and $K^0$
electromagnetic form factors,'' nucl-th/0005015.
%
\bibitem{mark} C.J.~Burden, C.D.~Roberts and M.J.~Thomson, Phys.\ Lett.\ {\bf
B 371} (1996) 163.
%
\bibitem{PDG} C.~Caso {\it et al.}, Eur.\ Phys.\ J.\ {\bf C 3} (1998) 1.
%
\bibitem{mike} M.A.~Pichowsky and T.-S.H.~Lee,
Phys.\ Rev.\  {\bf D 56} (1997) 1644; 
F.T.~Hawes and M.A.~Pichowsky, Phys.\ Rev.\ {\bf C 59} (1999) 1743.
%
\bibitem{FS} J.M.~Flynn and C.T.~Sachrajda, Preprint \heplat{9710057}.
%
\bibitem{richman} J. D. Richman and P. R. Burchat, Rev. Mod. Phys. {\bf 67},
893 (1995).
%
\bibitem{pirad} S. R. Amendolia {\it et al}., Nucl. Phys. B {\bf 277}, 168
(1986); 
C. J. Bebek, {\it et al}., Phys. Rev. D {\bf 13}, 25 (1976), 
{\it ibid} {\bf 17}, 1693 (1978);
%
\bibitem{cesr96} CLEO Coll. (J. E. Duboscq {\it et al}.),
Phys. Rev. Lett. {\bf 76}, 3899 (1996).
%
\bibitem{latt} UKQCD Coll. (D. R. Burford {\it et al}.), Nucl. Phys. B
{\bf 447}, 425 (1995).
%
\bibitem{gHsHpi} V. M. Belyaev, V. M. Braun, A. Khodjamirian and R. R\"uckl,
Phys. Rev. D {\bf 51}, 6177 (1995).
%
\bibitem{mrpion} P. Maris and C. D. Roberts, Phys. Rev. C {\bf 58}, 3659
(1998). 
%
\bibitem{kaondat} S. R. Amendolia, {\it et al}., Phys. Lett. B {\bf 178}, 435
(1986); W. Molzon, {\it et al}., Phys. Rev. Lett. {\bf 41}, 1213 (1978).
%
\bibitem{argus93} ARGUS Collaboration, Z. Phys. C {\bf 57}, 249 (1993).
%
\bibitem{a1b1} J.C.R.~Bloch, Yu.L.~Kalinovsky, C.D.~Roberts and S.M.~Schmidt,
Phys.\ Rev.\ {\bf D60} (1999) 111502.
%
\end{thebibliography}
\end{document}